\documentclass[sigconf,screen]{acmart}

\setcitestyle{numbers,sort&compress}
\usepackage{tikz}
\usepackage{comment}
\usetikzlibrary{positioning, arrows.meta}
\usepackage{booktabs}
\usepackage{pifont}
\usepackage{booktabs}
\usepackage{tabularx}
\usepackage{array}
\usepackage[most]{tcolorbox}
\tcbuselibrary{listings,breakable}
\usepackage{pgfplots}
\usepackage{pgfplotstable}
\usepackage{appendix}
\usepackage{xr-hyper}
\usepackage{placeins}

\AtBeginDocument{%
  }

\setcopyright{none}

\renewcommand\footnotetextcopyrightpermission[1]{}
\acmDOI{}
\acmISBN{}
\acmPrice{}
\acmConference[FSE '26]%
  {The 34th ACM Symposium on the Foundations of Software Engineering}%
  {July 5--9, 2026}%
  {Montreal, Canada}
\usepackage{fancyhdr}

\fancypagestyle{firstpageaccepted}{
  \fancyhf{}
  \fancyhead[C]{\small\itshape Accepted at ACM FSE 2026}
  
}
\acmBooktitle{Companion Proceedings of the 34th ACM Symposium on the Foundations of Software Engineering (FSE '26), July 5--9, 2026, Montreal, Canada}

\acmISBN{978-1-4503-XXXX-X/2018/06}

\pgfplotsset{compat=1.18} 
\begin{document}
\title{LACY: Simulating Expert Mentoring for Software Onboarding with Code Tours}
\author{Zeynep Begüm Kara}
\authornote{Both authors contributed equally to this work.}
\affiliation{%
  \institution{Bilkent University}
  \city{Ankara}
  \country{Turkey}
}
\email{begum.kara@alumni.bilkent.edu.tr}

\author{Aytekin İsmail}
\authornotemark[1]
\affiliation{%
  \institution{Bilkent University}
  \city{Ankara}
  \country{Turkey}
}
\email{aytekin.ismail@bilkent.edu.tr}

\author{Ece Ateş}
\affiliation{%
  \institution{Bilkent University}
  \city{Ankara}
  \country{Turkey}
}
\email{ece.ates@alumni.bilkent.edu.tr}

\author{İzgi Nur Tamcı}
\affiliation{%
  \institution{Bilkent University}
  \city{Ankara}
  \country{Turkey}
}
\email{izgi.tamci@alumni.bilkent.edu.tr}

\author{Zehra İyigün}
\affiliation{%
  \institution{Bilkent University}
  \city{Ankara}
  \country{Turkey}
}
\email{zehra.iyigun@alumni.bilkent.edu.tr}

\author{Selin Şirin Aslangül}
\affiliation{%
  \institution{Beko}
  \city{İstanbul}
  \country{Turkey}
}
\email{selinsirin.aslangul@beko.com}

\author{Ömercan Devran}
\affiliation{%
  \institution{Beko}
  \city{İstanbul}
  \country{Turkey}
}
\email{omercan.devran@beko.com}

\author{Baykal Mehmet Uçar}
\affiliation{%
  \institution{Beko}
  \city{İstanbul}
  \country{Turkey}
}
\email{baykal.ucar@beko.com}

\author{Eray Tüzün}
\affiliation{%
  \institution{Bilkent University}
  \city{Ankara}
  \country{Turkey}
}
\email{eraytuzun@cs.bilkent.edu.tr}

\begin{abstract}

Every software organization faces the onboarding challenge: helping newcomers navigate complex codebases, compensate for insufficient documentation, and comprehend code they did not author. Expert walkthroughs are among the most effective forms of support, yet they are expensive, repetitive, and do not scale. We present Lacy, a hybrid human-AI onboarding system that captures expert mentoring in reusable code tours—to our knowledge, the first hybrid approach combining AI-generated content with expert curation in code tours. Our design is grounded in requirements derived from 20+ meetings, surveys, and interviews across a year-long industry partnership with Beko. Supporting features include Voice-to-Tour capture, comprehension quizzes, podcasts, and a dashboard. We deployed Lacy on Beko's production environment and conducted a controlled study on a legacy finance system (30K+ LOC). Learners using expert-guided tours achieved 83\% quiz scores versus 57\% for AI-only tours, preferred tours over traditional self-study, and reported they would need fewer expert consultations. Experts found tour creation less burdensome than live walkthroughs. Beko has since adopted Lacy for organizational onboarding, and we release our code and study instruments as a replication package.

\end{abstract}

\renewcommand{\shortauthors}{Kara, İsmail, Ateş, Tamcı, İyigün, Aslangül, Devran, Uçar and Tüzün}


\maketitle
\thispagestyle{firstpageaccepted}

\section{Introduction}
Understanding code you did not write is one of the most challenging tasks in software engineering. Every software organization faces this challenge during onboarding, the process by which newcomers adjust to their new environment and learn the skills necessary to fulfill their roles to function effectively as members of an organization \cite{SaksUggerslevFassina2007}. Technical onboarding exposes newcomers to software architecture, development workflows, and coding standards \cite{Padoan2024OnboardingViz}, yet remains non-trivial due to high code complexity, lack of documentation, and difficulty getting help from a mentor or a senior developer \cite{BalaliEtAl2018}. Hence, a complete onboarding process for newcomers can be expected to take around three to six months, depending on the role and organizational practices  \cite{Brittoetal2018, Rollag2005getting}. Among the strategies organizations employ, such as courses, bootcamps, and mentorship \cite{santos2025software}, mentorship stands out as one of the most effective and desired forms of support \cite{ju2021case}. Expert walkthroughs support onboarding by transferring tacit knowledge, accelerating codebase navigation and workflow or tool mastery \cite{ryan2009development, dagenais2010moving}.

However, expert mentoring is inherently one-to-one and therefore difficult to scale. Since the knowledge conveyed in codebase walkthroughs often vanishes once the session ends, experts can find themselves repeating the same explanations for each new hire. If designed poorly, mentorship tasks reduce expert productivity and cost organizations valuable resources and talents \cite{Brittoetal2018, ju2021case,han2024understandingnewcomersonboarding, gao2025understandingcodebaselikeprofessional}. So, although expert guidance is highly desirable for newcomers, it remains inefficient for organizations to provide at scale.

Although tools such as GitHub Copilot\footnote{\url{https://github.com/features/copilot}} or Cursor\footnote{\url{https://cursor.sh}} can support newcomers in understanding individual code fragments, they struggle with holistic understanding due to complex interdependencies, custom algorithms unseen in training data, and limited training on comprehension tasks \cite{gu2025challengespathsaisoftware, eibl2025exploringchallengesopportunitiesaiassisted}. However, during onboarding, holistic understanding is commonly prioritized over implementation details \cite{ko2007information}. Newcomers need design decisions, business logic, and high-level overviews to establish context and know where to start \cite{dagenais2010moving}. Therefore, expert guidance remains essential.
 
Prior work on code tours \cite{codetour2022} and LLM-assisted tour generation \cite{Balfroid2024LLMGeneratedCodeTours, Balfroid2025LocalLLM} shows code tours can serve as useful entry points for understanding codebases. However, existing approaches either require significant manual effort to create or rely on fully automated generation that lacks expert insight. Neither captures the expert-newcomer relationship that makes mentoring valuable, nor addresses how code tours can be integrated into industrial workflows. 

We identify this gap and propose Lacy, an onboarding system that captures expert mentoring and encodes it in reusable code tours, simulating an expert walking through the codebase alongside the learner. Lacy combines AI assistance, which accelerates tour creation and explains generic code patterns, with expert input that contributes design rationale, organizational context, and judgment about what matters. Supporting features include Voice-to-Tour capture, comprehension quizzes, podcasts, Q\&A channels, and an expert dashboard. By preserving senior developers' knowledge in IDE-integrated tours, Lacy transforms ephemeral walkthroughs into scalable, reusable onboarding assets that mimic experts explaining code while navigating sequentially through relevant files.

To evaluate Lacy in an industrial setting, we partnered with Beko\footnote{\url{https://beko.com}}, a multinational home-appliance manufacturer with 50K employees. Through 20+ meetings with stakeholders we analyzed practices and onboarding challenges. We deployed Lacy on Beko's production environment and conducted a controlled study on a legacy codebase. Our findings indicate that expert-guided tours improve learner comprehension (83\% vs 57\% for AI-only in the quiz and 79\% and 76.8\% for AI-only in expert evaluations), reduce expert burden compared to live walkthroughs, and integrate effectively into existing workflows. Beko has since adopted Lacy to systematize its onboarding process and preserve expert knowledge across the organization. We define the following research questions (RQs) to assess our proposed system:

\begin{itemize}
    \item \textbf{RQ1:} How effectively does Lacy support experts in onboarding newcomers to unfamiliar codebases?
    \item \textbf{RQ2:} How effectively does Lacy support newcomers in understanding unfamiliar codebases?
    \item \textbf{RQ3:} How effectively does Lacy address organizational onboarding needs and integrate into existing workflows?
\end{itemize}

Our contributions are fourfold: (i) We derive requirements for software onboarding tools through a year-long industry partnership involving surveys, interviews, and co-design sessions. (ii) We propose Lacy, a software onboarding system that addresses challenges faced by both experts and learners during onboarding. (iii) We propose AI-assisted code tours to simulate expert mentoring and demonstrate their feasibility in an industrial setting. (iv) We deploy Lacy in Beko's production environment and release our code in the replication package\footnote{\url{https://figshare.com/s/6a261d3382b116d8494f}}.

\section{Industry Context \& Onboarding Challenges}
\label{subsec:Onboarding Challenges at Beko}
We partnered with Beko for over a year to understand onboarding challenges and co-design solutions. The collaboration brought together product and engineering leadership, engineering leads, senior developers, and new hires through 20+ meetings, surveys, and interviews across three phases: problem discovery, system design, and evaluation.

\textbf{Current Practices.} Beko's onboarding process is semi-structured, relying on informal one-on-one mentorship. Although the onboarding experience depends on team dynamics, role responsibilities, and individual differences, new hires at Beko typically require guidance for approximately the first two months before becoming independently productive. Team members, especially team leads, support new hires with a mix of quick meetings, question answering on Microsoft Teams, and occasional scheduled check-ins while the biggest challenge they face is limited time dedicated for these tasks. Also, mentorship and expert-led code walkthroughs are common and highly valued, as they provide practical guidance. 

\begin{table*}[!htbp]
\centering
\caption{Industry requirements for an effective onboarding system}
\label{tab:requirements}
\small
\begin{tabular}{@{}p{5cm}p{7.5cm}p{4.5cm}@{}}
\toprule
\textbf{Requirement} & \textbf{Rationale} & \textbf{Addressed By} \\
\midrule
\textbf{R1}: Capture and scale expert guidance & Walkthroughs are effective but expensive, ephemeral, and require repetition for each newcomer & Code Tour Structure, Podcasts \\
\addlinespace
\textbf{R2}: Reduce expert authoring burden & Mentors lack time; context-switching is cognitively exhausting & AI-Assisted Guided Tours, Voice-to-Tour \\
\addlinespace
\textbf{R3}: Preserve and facilitate the transfer of tacit knowledge & Design rationale and conventions exist only in experts' minds; LLMs cannot infer organizational context without explicit documentation & Guided Tours \\
\addlinespace
\textbf{R4}: Enable self-directed learning & Limited expert availability; social friction; autonomy to explore at their own pace & Exploratory Tours, Note Taking \\
\addlinespace
\textbf{R5}: Provide structured entry points & Learners feel ``dropped in the middle'' with no starting point & Code Tour Structure \\
\addlinespace
\textbf{R6}: Enable comprehension assessment & Current assessment is informal with no systematic verification & Quizzes, Dashboard \\
\addlinespace
\textbf{R7}: Support asynchronous communication & Q\&A can interrupt work; message-based questions lack context & Q\&A, Feedback \\
\addlinespace
\textbf{R8}: Record organic mentoring as it happens & Experts explain code to colleagues daily but this knowledge vanishes; it should be captured in real-time rather than recreated later & Voice-to-Tour \\
\bottomrule
\end{tabular}
\end{table*}

\textbf{Challenges.} Although code walkthroughs are effective, our interviews revealed they impose significant costs on both sides. Senior developers reported spending 10–20 hours per new hire, often repeating the same explanations: \textit{``Alongside your regular work, onboarding and mentoring someone is a challenging mental effort—if teams change frequently, we do it many times for many people''.} Scheduling coordination and ad-hoc questions further fragment their focused work time.  Newcomers described needing help frequently but hesitated to ask due to reluctance to interrupt busy colleagues, uncertainty about what to ask, and confusion about whom to approach, which leads many to reach out only when absolutely necessary. These dynamics are compounded by knowledge concentration. Interviewees confirmed that critical parts of the codebase are understood by only one or two people and that, when those individuals leave, their knowledge leaves with them: \textit{``People who knew the business logic and design decisions left and we are left with nothing and no guide for others''.} the Director of IT Products \& Services Development noted. This \textit{bus factor} \cite{avelino2016novel, jabrayilzade2022bus} problem underscores the need to capture expert knowledge before it is lost.

\textbf{Prior Attempts.} Beko has attempted to address these challenges through traditional documentation and AI tools, but stakeholders agreed that neither suffices. Developers reported minimal documentation at Beko and noted that existing documentation formats do not capture mentor walkthroughs, since such formats are generally prepared for people who know what they are looking for. Even when documentation exists, it quickly becomes outdated: \textit{``Often only the code is updated; therefore, we need people who understand the code implementation, in this case experts, to share their knowledge''.}. Learners are encouraged to use AI-powered tools such as  GitHub Copilot or Cursor, which offer valuable support to explain individual code snippets, but complement rather than replace the high-level guidance experts provide about architecture, team practices, and design decisions. The IT Director emphasized that AI-powered tools do not capture tacit knowledge, which matters more than implementation details during onboarding: \textit{``Only about 20\% of the code handles the critical business logic, and AI cannot tell you which 20\%. Relying solely on AI is insufficient for onboarding because mentors provide context and design decisions not inferable from code''}. Expert guidance remains the most valuable onboarding resource at Beko, yet it is ephemeral and does not scale with Beko's onboarding demand.

\textbf{Design Opportunity.} These challenges present a design opportunity. Onboarding requires newcomers to learn many aspects of a system, among which undocumented and tacit knowledge is particularly critical but difficult to transfer through traditional documentation or extract from code alone. Mentoring remains one of the few effective ways to convey this expert knowledge. Rather than viewing this dependency as a bottleneck, we see mentoring as an ideal moment to capture expert insights that would otherwise stay implicit. Our design aims to encode this guidance in reusable code tours that simulate expert mentoring for future references and scale beyond one-on-one interactions, reducing burden on seniors while fostering a culture of knowledge sharing that benefits both experts and learners. From these observations, we derived a set of design requirements for Lacy, presented in Table~\ref{tab:requirements}.

\section{Related Work} \label{sec:Related Work}
\subsection{Documentation and Mentoring}

Documentation is fundamental to software engineering, supporting maintenance, collaboration, and preservation of organizational memory ~\cite{arya2024people}. During onboarding, developers must quickly learn code semantics, API details, and domain-specific concepts, yet the relevant information is often insufficient and dispersed across multiple sources; newcomers are expected to get accustomed to the company standards from READMEs, wikis, and inline comments ~\cite{nam2024usingllmhelpcode}. However, documentation becomes outdated as projects evolve, is inconsistent in quality, and is time-consuming to create ~\cite{Tan_2025, aghajani2020software}. Experts are often reluctant to document because it competes with primary coding responsibilities~\cite{arya2024people}. Instead, they prefer synchronous communication such as quick meetings or chat messages since it feels faster and more directly helpful~\cite{storey2014r}, leaving critical knowledge undocumented.

Even setting aside these challenges, mentoring remains a well-established onboarding practice~\cite{britto2020evaluating, santos2025software}. Studies consistently highlight mentoring-based approaches as highly effective in onboarding processes, as they are also one of the first forms of support newcomers turn to when dealing with an unknown codebase~\cite{bexell2024firstbug, rodeghero2021please, dagenais2010moving}. This includes undocumented information that exists only in experts' minds~\cite{ryan2013acquiring} such as the intent behind code already written~\cite{ko2007information}, design rationale, historical context \cite{latoza2006maintaining}, and known pitfalls~\cite{gregory2022onboarding}. Also, mentoring can transfer tacit knowledge that documentation fundamentally cannot capture~\cite{nonaka2009knowledge, bjornson2008knowledge}. Mentoring provides newcomers with a guiding narrative about where to start, how components connect, and what to prioritize~\cite{dagenais2010moving}, and it helps them think through “hard-to-answer questions about code”~\cite{latoza2010hard}.

However, mentoring does not scale: it must be repeated for each newcomer, it requires effort for experts~\cite{dagenais2010moving}, mentors are often occupied with other responsibilities~\cite{begel2008struggles}, and knowledge is lost when experts leave~\cite{robillard2021turnover, avelino2016novel}. As Padoan et al.~\cite{Padoan2024OnboardingViz} emphasize: ``We have to frequently allocate some team members to provide more support to the new team member, and it incurs delays. While a wide range of documentation tools is available \cite{misback2025codetationsintelligentpersistentnotes}, to our knowledge no comparable artifact exists for capturing mentoring. Despite their demonstrated effectiveness, high demand, and intuitive nature, expert walkthroughs remain fundamentally ephemeral, unable to be preserved or reused.

\subsection{Code Tours}
The concept of guided code walkthroughs has roots in industry practice. Johnson and Senges document Google's internal "Codewalks"—"a mechanism for tracing through examples of technology use in the real code base" where "a viewer takes the Noogler through real source code step-by-step". This approach was developed because tutorials do not really prepare the engineers to perform 'in the wild'" \cite{johnson2010learning}. The Go programming language's documentation defines a codewalk as "a guided tour through a piece of code" consisting of "a sequence of steps, each typically explaining a highlighted section of code" \cite{go-codewalk}.
Microsoft's CodeTour VSCode extension\footnote{\url{https://marketplace.visualstudio.com/items?itemName=vsls-contrib.codetour}. As of January 2026, it has over 422,640 installs and a 5-star rating from 21 reviewers.} further popularized this approach. Taylor and Clarke~\cite{codetour2022} define a code tour as a linked list of nodes where each node points to a file location and provides an explanation that guides readers through a workflow or design rationale. Code tours can help developers to see how various parts of the codebase relate to each other in a specific context. It is argued that a well-written code tour can serve as a valuable starting point for developers learning a new codebase~\cite{codetour2022}. However, creating a well-written code tour can be time-consuming, as it requires careful selection of relevant code locations, clear explanations, and a logical sequence that reflects the workflow. As shown in Table ~\ref{tab:lacy_vs_codetour}, Lacy introduces more automation to create and share code tours and enhances user interaction.

Balfroid et al.~\cite{Balfroid2024LLMGeneratedCodeTours} initiated work on LLM-assisted code tour generation, arguing that automated tours could reduce the burden on senior developers. Their approach focuses on debugging scenarios, generating tours by capturing stack traces from failed tests,  extracting code snippets with CodeQL, and producing explanations using GPT-3.5. In later work, Balfroid~\cite{Balfroid2025LocalLLM} investigates locally runnable LLMs for the same purpose and argues that automated code tour generation remains underexplored. The author outlines key research 
directions, such as selecting relevant code segments, generating documentation, and evaluating quality, and emphasizes that segment selection in large repositories is a core challenge.

We share this view on both the potential of code tours and the challenges of automated tour generation. However, rather than fully automated approaches focusing on debugging scenarios, we propose a hybrid human-AI approach that combines LLM-generated content with expert refinement. These approaches are defined in Table~\ref{tab:lacy_vs_codetour} as fully automated and AI-assisted tour generation. Our approach aims to generalize the use of AI in creating code tours that simulate expert guidance and to outline a framework for adopting code tours in industry.

\begin{table}[t]
  \caption{Comparison of \textit{Lacy}, \textit{CodeTour}, \& \textit{Balfroid et al.}}
  \Description{Comparison table showing which features are supported by Lacy, CodeTour, and Balfroid et al.'s approach. A checkmark indicates support, a cross indicates lack of support.}
  \label{tab:lacy_vs_codetour}
  \centering
  \begin{tabular}{p{4.2cm}ccc}
    \toprule
    \textbf{Feature} & \textbf{Lacy} & \textbf{CodeTour} & \textbf{Balfroid} \\
    \midrule
    Navigation with tours             & \ding{51} & \ding{51} & \ding{51} \\
    Manual tour generation            & \ding{51} & \ding{51} & \ding{55} \\
    AI-assisted tour generation       & \ding{51} & \ding{55} & \ding{55} \\
    Fully automated tour generation   & \ding{51} & \ding{55} & \ding{51} \\
    Assigning tours to learners       & \ding{51} & \ding{51} & \ding{55} \\
    Voice-to-Tour generation          & \ding{51} & \ding{55} & \ding{55} \\
    Quizzes                           & \ding{51} & \ding{55} & \ding{55} \\
    Podcasts                          & \ding{51} & \ding{55} & \ding{55} \\
    Dashboard                         & \ding{51} & \ding{55} & \ding{55} \\
    Industrial deployment             & \ding{51} & \ding{55} & \ding{55} \\
    Onboarding focus                  & \ding{51} & \ding{55} & \ding{55} \\
    Debugging focus                   & \ding{55} & \ding{55} & \ding{51} \\
    Linking tours to each other       & \ding{55} & \ding{51} & \ding{55} \\
    \bottomrule
  \end{tabular}
\end{table}

\subsection{AI-Assisted Code Understanding}
AI-powered tools such as GitHub Copilot and 
Cursor are increasingly used for code understanding. Khojah 
et al.~\cite{khojah2024codegenerationobservationalstudy} found that 62\% of 
ChatGPT use in software engineering involves consultations. Developers seek 
guidance rather than concrete solutions, effectively treating LLMs as ``virtual colleagues ~\cite{khojah2024codegenerationobservationalstudy}''. However, LLMs struggle with holistic codebase understanding: Gu et al.~\cite{gu2025challengespathsaisoftware} show that complex interdependencies, custom algorithms, and limited context windows prevent LLMs from reasoning about how components connect across a system. Moreover, AI effectiveness depends on prior familiarity. Kumar et 
al.~\cite{kumar2025sharptoolsdeveloperswield} found that experienced 
contributors provided expert-level insights 70\% of the time when using AI tools, compared to 48\% for less experienced participants. This suggests AI tools benefit developers who already understand the codebase more than newcomers who need to learn it.

Several frameworks have been proposed to use AI to mimic a senior developer. Ionescu et al.~\cite{ionescu2025multiagentonboardingassistantbased} introduced \textit{Onboarding Buddy}, an IntelliJ plugin that combines retrieval-augmented generation and automated chain-of-thought with an LLM interface to offer an end-to-end onboarding system, minimizing human interaction. Tan et al.~\cite{Tan_2025} proposed OSSerCopilot, a platform-integrated interface envisioned as a future AI mentor for newcomers. However, these approaches operate as reactive tools, answering questions when asked rather than providing structured learning paths that guide newcomers through the codebase step by step. Furthermore, by minimizing human involvement, they eliminate the expert-newcomer communication.

To the best of our knowledge, prior work either offers static, manually-authored code tours or investigates fully automated tour generation, but does not combine AI assistance with expert curation to simulate mentorship at scale. No published feasibility study examines how code tours integrate into organizational onboarding workflows or evaluates their effectiveness. Moreover, existing tools lack the supporting mechanisms—comprehension quizzes, asynchronous Q\&A, progress tracking, and alternative formats like podcasts—needed to simulate the experience of an expert sitting next to you, pointing at relevant code, and walking you through the codebase.

\begin{figure*}[t]
  \centering
  \includegraphics[width=\textwidth]{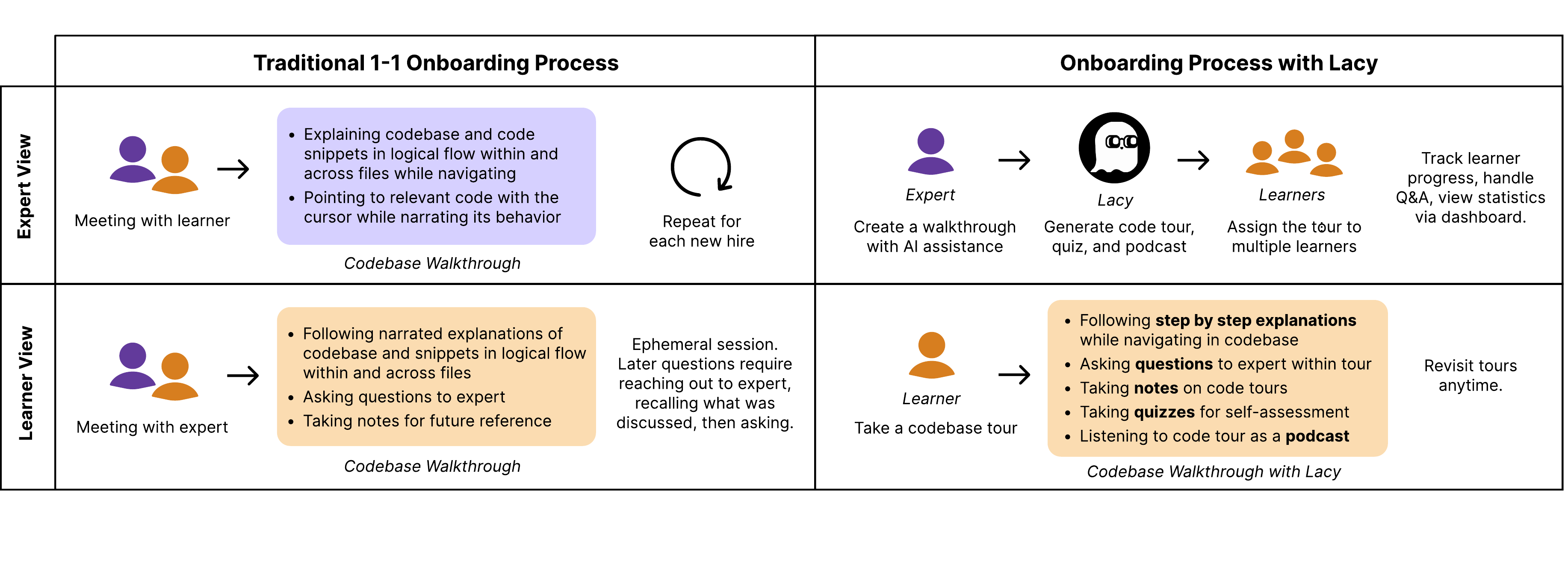}
  \caption{Expert and learner views of traditional 1-1 onboarding versus onboarding with Lacy. Lacy preserves the core learning experience while adding benefits for both: easier walkthrough creation for experts, richer learning resources for learners.
  }
  \label{fig:trad_vs_lacy}
\end{figure*}

\section{Industry-Driven Design and Features} \label{sec:Industry-Driven Design and Features}
The challenges identified in Section~\ref{subsec:Onboarding Challenges at Beko} point to a central tension: expert walkthroughs are highly effective but very inefficient. Lacy solves this by capturing walkthrough-style guidance with code tours. Code tours are step-by-step, code-anchored explanations that learners can follow independently, preserving the narrative structure and contextual grounding that traditional documentation lacks, the types of code tours users can create are detailed in Table \ref{tab:tourtables}. Lacy is implemented as a VS Code extension, integrating tours directly into the development environment where explanations and code can be viewed together. Throughout the system, we distinguish two user roles:

\textbf{Codebase experts:} Developers with deep knowledge of a codebase, responsible for creating tours and supporting newcomers.

\textbf{Codebase learners:} Developers seeking to understand an unfamiliar codebase, whether as new hires, project transfers, or maintainers taking over new responsibilities.

These roles are context-dependent; the same developer may be an expert on one project and a learner on another. Our design emphasizes asynchronicity and reusability, enabling experts to invest effort once while multiple learners benefit independently. All features are designed to support asynchronous communication (R7), eliminating the need for synchronous scheduling between experts and learners, unlike traditional 1-1 mentoring sessions. The perspectives of both user types within each onboarding approach are summarized in Figure~\ref{fig:trad_vs_lacy}. 

\begin{table}[!b]
\caption{Code Tour types in Lacy}
\label{tab:tourtables}
\renewcommand{\arraystretch}{1.15}
\setlength{\tabcolsep}{6pt}

\hfill 
\begin{tabularx}{\columnwidth}{
  >{\raggedright\arraybackslash}p{0.15\columnwidth}
  >{\raggedright\arraybackslash}p{0.48\columnwidth}
  >{\raggedright\arraybackslash}X
}
\toprule
\textbf{Type} & \textbf{Authorship \& Audience} & \textbf{Process} \\
\midrule
Guided & Experts using LLM for learners & Manual, AI-assisted, Voice-to-Tour \\
Exploratory & Learners using LLM for self-use & AI-generated \\
\bottomrule
\end{tabularx}
\end{table}

\subsection{Code Tours as Guides}
\begin{figure}
\centering
\includegraphics[width=1\linewidth]{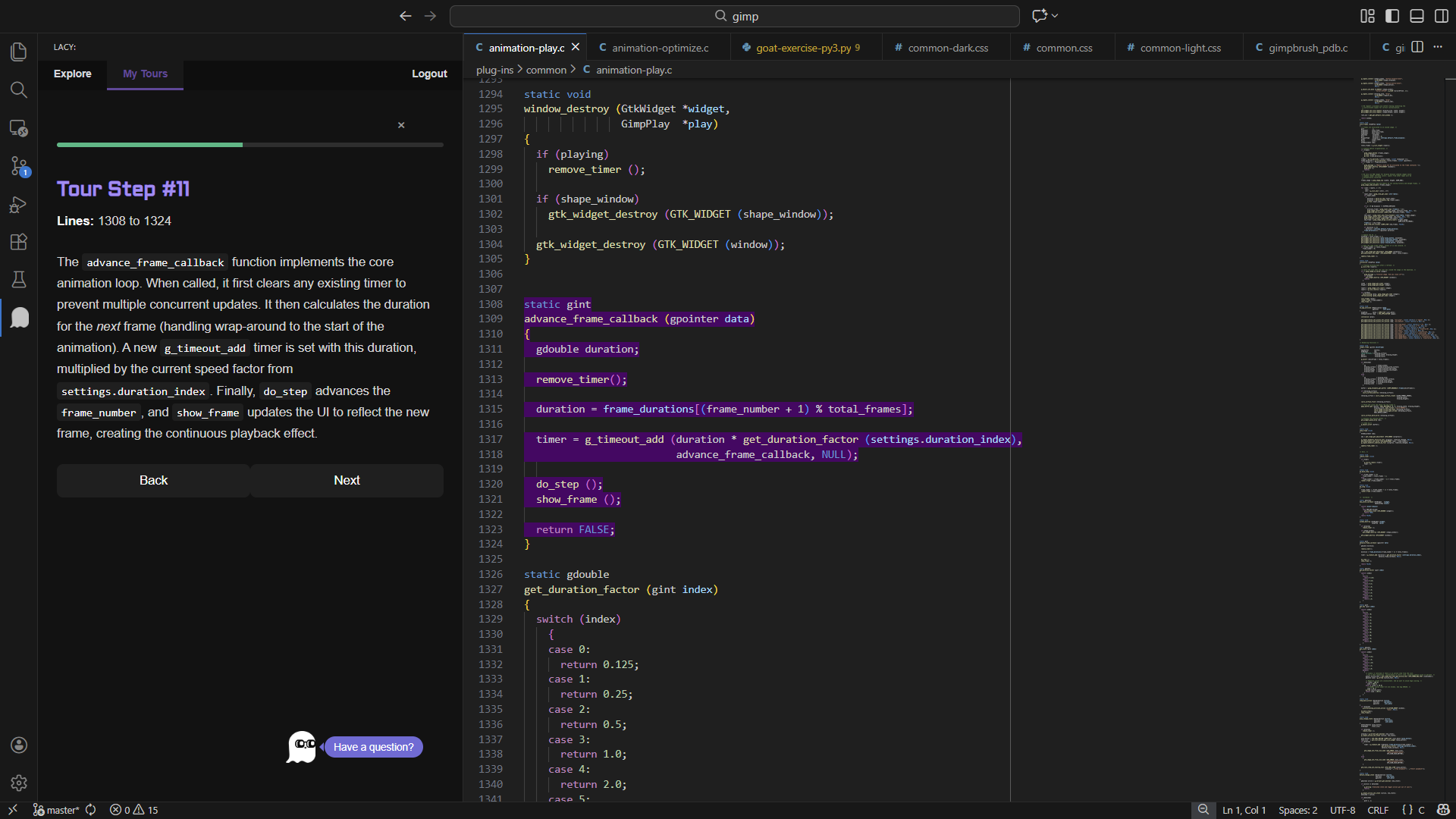}
\caption{Code Tour User Interface}
\label{fig:tour-step}
\end{figure}

Code Tour is the most essential feature of Lacy where code explanations are given side by side with the files and the relevant code segments are highlighted as shown in Figure \ref{fig:tour-step}. With them, Lacy aims to replicate the one-to-one interaction that would be needed during the traditional mentoring sessions. Moreover, one of the problems we had identified in other tools is the lack of opportunities for learners to ask questions or give any feedback to the tour’s creators through the same UI. Our feedback and ask-question features were designed to address this gap. Finally, Lacy’s codebase tours can be complemented with quizzes related to the generated tour or accompanied by a conversational podcast that presents the tour’s content in an alternative medium.

In Lacy, users can create tours for different purposes, each designed with a specific real-life scenario in mind. There are two fundamental tour types: 
\textbf{guided tours} are created by experts to detail a certain segment of the codebase for the learners, and \textbf{exploratory tours} can be created by all users to gain insight about the codebase with AI-generated tour content.

\subsubsection{Guided Tours}
\begin{figure}
\centering
\includegraphics[width=1\linewidth]{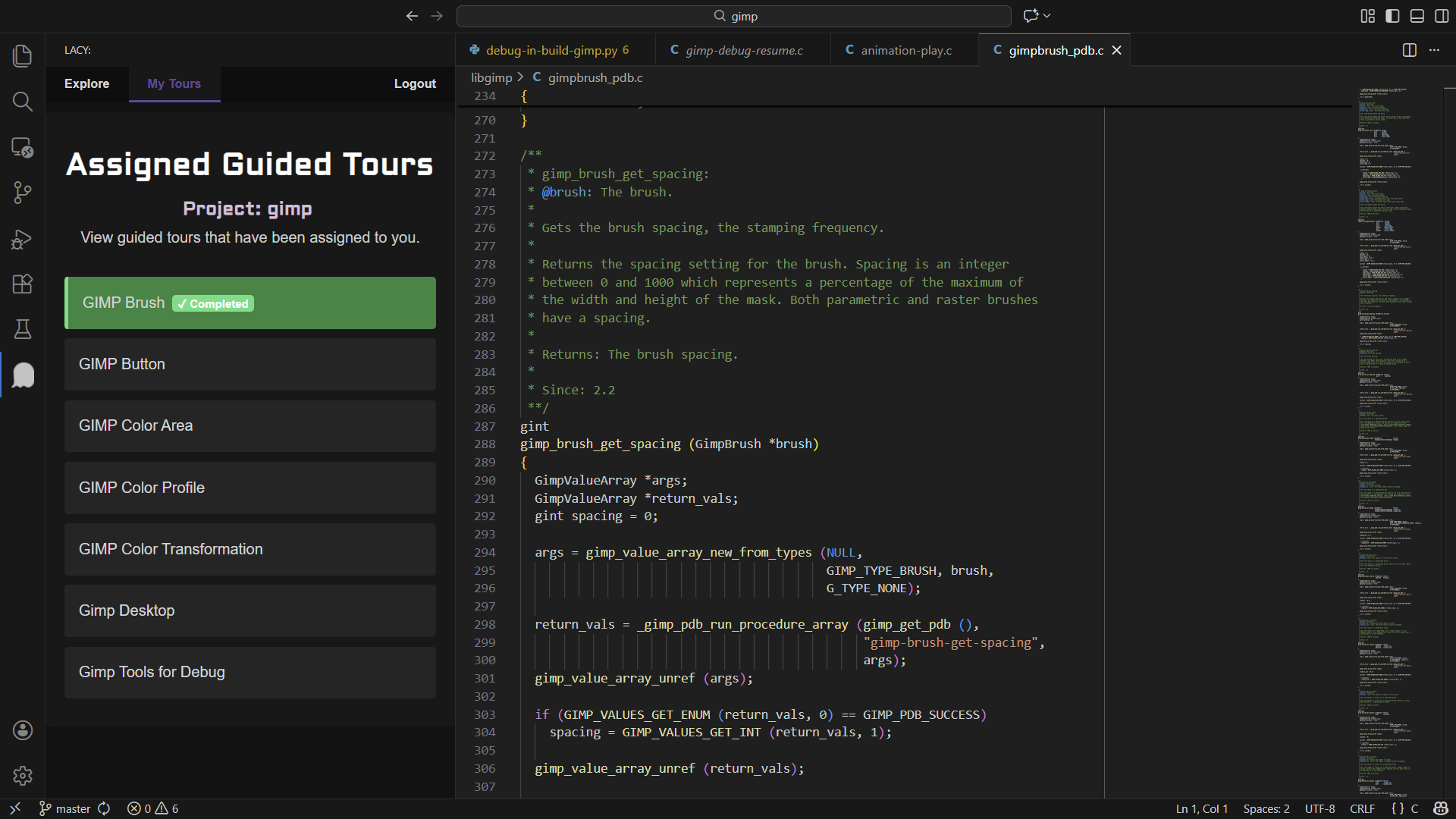}
\caption{Assigned Tours User Interface}
\label{fig:assigned-tour}
\end{figure}

Senior developer guidance and LLM explanations are two of the most valuable forms of support available to new hires during technical onboarding. Because each approach has clear limitations when used in isolation, Lacy prioritizes guided tours that combine their strengths and mitigate their drawbacks. Guided tours can only be created by the expert users of a given codebase and are automatically assigned to the learners. The list of assigned tours will be visible to learners as shown in Figure \ref{fig:assigned-tour}. To create a guided tour, experts define the intent and scope of a tour, the LLM produces an initial draft, and the expert then refines it before it is shared with learners.

Our goal in allowing experts to edit their guided tours was to enable them to correct any mistakes that the LLM might make. At the same time, this editing allows experts to improve the tour by adding any codebase-specific knowledge that LLMs are likely to miss, while maintaining the speed and efficiency benefits of LLM content generation. Hence, the editing process is essential for meeting requirements R1 and R5, as without expert input, any attempt to capture their guidance and provide an accurate starting point for understanding the codebase would remain incomplete. Also, we seek to address the repetitiveness of the mentoring task by storing context-specific knowledge in an accessible medium to meet the requirements R2 and R3. This approach aims to minimize the need for mentors to repeatedly explain the same practices to each new hire and, hence, facilitate the onboarding process. 

The manual tour feature was added to Lacy to complement the guided tour and provide a comprehensive tour creation experience for codebase experts. In manual tour creation, experts are given full control over all aspects of the tour. For each step, the expert needs to specify the relevant file, line interval, and corresponding code explanation. As they are authored completely by experts, manual tours meet the same requirements as AI-assisted tours except R2; specifically, they satisfy R1, R3, and R5. While we expect manual tours to be created less frequently than AI-assisted tours, we consider this feature essential for covering all facets of the code tour generation process.

\subsubsection{Exploratory Tours}
Exploratory tour feature’s main purpose is to give users the freedom to explore the codebase through fully AI-generated tours, without relying on expert availability. They allow all users to explore any part of the codebase without having to wait for the creation of a guided tour. In doing so, we aim to encourage users to develop a comprehensive understanding of the codebase on their own initiative, thus meeting requirement R4. However, since exploratory tours lack senior developer input, they carry the risk of providing a superficial or incomplete explanation of the inquired section. For that reason, exploratory tours are positioned as an approachable starting point for codebase orientation, while guided tours remain as the primary source for context-aware and reliable explanations.

\subsection{Supporting Features}
Beyond code tours, Lacy includes features that simulate other aspects of expert guidance during traditional onboarding.

\subsubsection{Voice-to-Tour}
During co-design sessions, experts noted that valuable explanations happen organically during explaining codebase to newcomer or during pair programming but are never captured. In contrast, traditional documentation requires dedicated time and effort outside day-to-day development. Lacy's Voice-to-Tour feature addresses this tension by capturing natural mentoring interactions and converting them into structured, reusable code tours that simulate expert guidance for future learners.

 When an expert is explaining code to a colleague and wants to preserve that explanation, they start a recording in extension. Lacy captures the interaction context, including which files are selected and the order in which topics are covered, with spoken narration. After the session, this content is sent to an LLM, which organizes it into a coherent code tour satisfying R1, R2, R3, R5 and R8.

This workflow preserves what code tours are meant to capture: conversational tone, coherent information flow, design rationale, and contextual insights. This workflow is possibly the most intuitive and faithful way to generate code tours, because it captures expert guidance in its natural setting. By converting naturally occurring mentoring interactions into structured tours, Lacy turns otherwise ephemeral explanations into reusable onboarding assets that can be replayed and referred later.

\subsubsection{Podcast}
To complement the visual code tour experience, Lacy includes a podcast feature that converts tour content into a conversational dialogue using text-to-speech synthesis. Instead of producing a single-speaker narration, Lacy generates a NotebookLM\footnote{\url{https://notebooklm.google/}}-style dialogue script between a simulated expert and learner, reflecting the dynamics of real mentoring sessions: the learner asks clarifying questions, and the expert responds with explanations and practical insights.

This feature supports different learning preferences and by allowing users to shape their onboarding journey according to their needs, satisfies the R4 requirement. While listening, learners can explore the codebase in parallel, following the expert’s design reasoning as they would during an in-person walkthrough. Combined with the tours, the podcasts provide a closer approximation of the interactive mentoring experience.

\subsubsection{Quizzes}
During in-person mentoring, experts naturally assess understanding through informal checks. These moments help identify knowledge gaps early and allow experts to adjust their explanations. Lacy's quiz feature replicates this by converting tour content into targeted comprehension questions. For learners, quizzes consolidate newly acquired knowledge and reinforce key concepts. For experts, quiz results indicate whether a tour is achieving its instructional goals and where explanations may need improvement.

When creating an AI-assisted guided tour, the LLM generates initial quiz questions based on the covered code. Experts then review and refine them, potentially adding domain-specific questions targeting common misconceptions, meeting requirements R1 and R3. Questions are linked to specific tour steps, allowing learners to revisit relevant code when reviewing incorrect answers. Results are displayed on the expert dashboard, providing visibility into individual and aggregate comprehension over time, satisfying R6.

\begin{figure*}
  \centering
  \includegraphics[width=\textwidth]{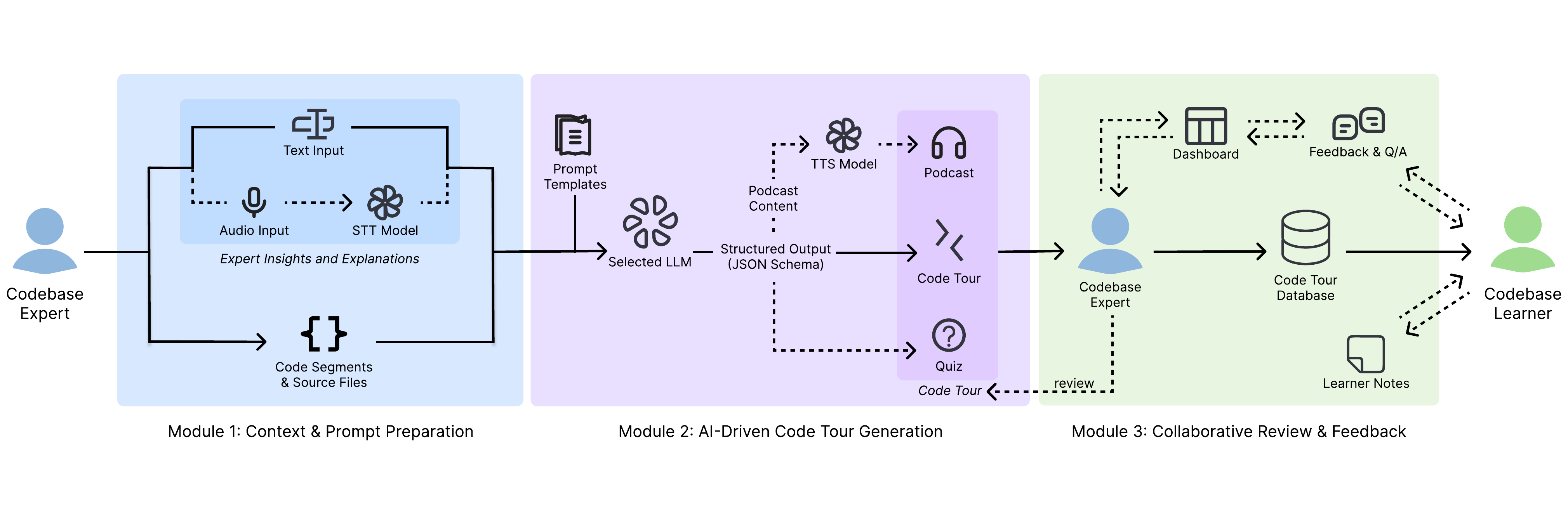}
  \caption{%
    \textbf{AI-assisted guided tour pipeline.}
    System architecture illustrating the three modules:
    (1)~Context \& Prompt Preparation,
    (2)~AI-Driven Code Tour Generation, and
    (3)~Collaborative Review, Feedback, \& Iteration.
    \textit{Solid lines} indicate the primary workflow between modules,
    while \textit{dashed lines} represent optional processes.%
  }
  \label{fig:pipeline}
\end{figure*}
\subsubsection{Note Taking, Q/A \& Feedback}
In effective mentoring, learners are not passive recipients—they take notes to anchor their understanding, ask questions when something is unclear, and signal when explanations succeed or fall short. Lacy supports this active engagement. While viewing any tour, learners can take notes on individual steps to record anything for future reference. These notes remain attached to the relevant tour step, keeping personal annotations organized and accessible during later review, and allow users to modify tours for their needs, satisfying R4.

Beyond note-taking, Lacy establishes a bidirectional channel between learners and experts. If a learner has a question about a specific step, they can submit it directly through the interface. The question appears on the expert dashboard, where it can be reviewed and answered asynchronously, simulating the natural Q\&A flow of in-person mentoring without requiring both parties to be available simultaneously, satisfying R7. Learners can also rate each tour and leave comments on its clarity and usefulness. This feedback signals to experts which tours are effective and which may need revision, creating an improvement loop where future learners benefit from progressively better guidance. Together, these features ensure that code tours support active learning and remain improvable.

\subsubsection{Expert Dashboard}
When expert guidance is delivered through code tours, learner progress becomes visible to both experts and management. Lacy's dashboard allows experts to monitor individual learners, including completion status and quiz performance, helping identify where learners struggle and enabling targeted support or tour revisions. Beyond individual progress, the dashboard displays aggregate statistics across teams and projects, identifying gaps in tour coverage. Patterns in exploratory tour usage can also signal unmet needs such as multiple learners generating tours on similar topics may indicate a gap that a guided tour could address.

The dashboard serves as the central hub for the asynchronous Q\&A workflow described earlier. Experts can review questions submitted on specific tour steps and respond directly; answers appear beneath the original question in the tour interface, closing the communication loop without synchronous interaction. Experts can also assign specific tours to learners and track completion throughout their team, making onboarding progress measurable and allowing managers to identify when newcomers are ready to move on or where additional support is needed, satisfying R6.

\subsection{System Design} \label{sec:System Design}
Lacy supports two types of code tours (Table~\ref{tab:tourtables}): 
\textit{Guided Tours} and \textit{Exploratory Tours}. 
In this section, we focus on the \textit{AI-assisted Guided Tour Pipeline}, 
which forms the core of Lacy and exemplifies collaboration between 
expert knowledge and AI-generation. 
Figure~\ref{fig:pipeline} illustrates this workflow's architecture and its three main modules.

\subsubsection{Module 1: Context \& Prompt Preparation}
When Lacy is executed as a VS~Code extension on an active project, it automatically retrieves tours associated with that workspace. Experts can provide input both through text and audio, supported by the integrated speech-to-text (STT) modules\footnote{\url{https://elevenlabs.io}}. The extension allows natural file and code-segment selection using cursor highlighting, enabling experts to define the scope of the tour generation process.

\subsubsection{Module 2: AI-Driven Code Tour Generation}
The collected context is formatted into predefined prompt templates for tour and podcast generation. Lacy supports multiple LLM configurations, allowing users to select a model for each generation task. The current deployment at Beko uses Google's Gemini-2.5-Flash \cite{comanici2025gemini}. When API access is restricted, local model endpoints can be utilized. The model outputs a structured JSON containing tour steps, explanations, and quiz items, which are rendered in an editable interface within the extension for expert revision. For podcast generation, the LLM also produces a two-person dialogue script simulating a conversation between an expert and a learner; this script is then converted to audio using ElevenLabs text-to-speech, (TTS) with distinct voices for each speaker.

\subsubsection{Module 3: Collaborative Review \& Feedback}
After generation, experts edit or approve tours before publication. Once finalized, the tour is stored in the company’s internal database and the tour is assigned to the selected learner group. Podcasts are also stored as binary large objects in the same database. Learners can access published tours directly from their workspace. Individual learner notes remain private, while learner engagement and performance metrics are displayed in the expert dashboard for iterative improvement and tracking.

\subsubsection{Other Tour Pipelines}
Beyond the AI-assisted Guided Tour Pipeline, Lacy supports two complementary pipelines offering different balances of human and AI contribution, depending on the onboarding context and authoring needs.

\textit{Manual Guided Tour Pipeline} grants full control over content and pedagogical intent with no AI involvement, encouraging the preservation of tacit knowledge that cannot be inferred from source code alone. This IDE integrated step by step explanations can serve as an alternative documentation method.

\textit{Exploratory Tour Pipeline}, in contrast, supports fully automated tour generation through the language model over selected files and code segments. It provides learners with an immediate entry point when no guided tours are available, helping users quickly orient themselves within unfamiliar codebases.

Together, these three pipelines establish a spectrum of human–AI involvement in tour creation, considering the trade-off between knowledge transfer depth and generation speed. This spectrum ranges from fully human-authored to fully AI-generated tours, allowing Lacy to adapt to diverse documentation and learning scenarios. Lacy thus embodies a realistic collaboration between humans and AI, bridging manual expertise with automation.

\section{Beko Case Study} \label{sec:case-study}
Lacy is deployed on Beko's DevOps environment as a VSCode extension, available to all developers across teams. Throughout deployment, we gathered continuous feedback from developers and decision-makers, informing iterative improvements. To complement this deployment experience with rigorous evaluation, we conducted two complementary activities: (1) a controlled experiment assessing learning outcomes, (2) semi-structured interviews with Director of Digital Products \& Services and Head of IT.

\textbf{Pilot Project.} For the controlled study, we selected \textit{Bankhet}, a legacy finance application written in VB.NET in 2007 that remains in active use today. Bankhet is a business critical project for Beko since company payments are transferred to banks through this platform. When Bankhet is not operational, accounting personnel must create the electronic funds transfer manually, obtain the required approval, and pay for the transfer fees. The codebase is 30K+ lines of code and requires ongoing maintenance. Critically, the only two engineers with substantial institutional knowledge of Bankhet have since moved into managerial positions, which materially limits their availability to support the legacy system. Bankhet was selected precisely because it exemplifies the onboarding challenges Lacy aims to address: scarce documentation, limited expert availability, and high dependency on tacit knowledge.

\textbf{Participants.} We recruited seven participants from Beko: two domain experts and five developers onboarding onto Bankhet. Since institutional knowledge of Bankhet was concentrated in just two engineers, we recruited both as the study’s domain experts. The learners were five potential maintainers of the Bankhet project, consisting of three new hires and two cross-project developers. Crucially, all participants were real stakeholders with genuine responsibilities for Bankhet (not external volunteers) which ensured ecological validity by capturing authentic motivations, constraints, and learning behaviors. Both experts had extensive experience (16 and 22 years) and deep Bankhet knowledge, while learners averaged 4.6 years of programming experience with little to no prior experience working with Bankhet.

\textbf{Participant Background.} To contextualize our findings, the following results are drawn from pre-study surveys of learners and experts (LS-Pre, ES-Pre)\footnote{Survey naming: LS = Learner Survey, ES = Expert Survey}. 

From the learner perspective, onboarding support had been informal: 80\% had informal mentors, and 20\% had no mentor at all. Learners received support in multiple formats: 60\% of participants received code walkthroughs, 60\% received mentor/buddy assistance, and 40\% received starter tasks. Notably, no learners received architecture sessions or a documentation guide. Time to productivity varied considerably: 20\% achieved it in 2--4 weeks, 40\% in 1--2 months, and 40\% required 5+ months. Learners reported moderate comfort asking questions (M=2.8/5), with primary barriers being reluctant to interrupt colleagues (80\%) and uncertainty about what to ask (80\%). When learning unfamiliar code, they relied heavily on asking colleagues (M=4.2/5) and AI tools (M=4.0/5), followed by scheduling expert walkthroughs (M=3.8/5). Learners ranked ``the code was easier to understand'' as the top benefit of 1-on-1 expert explanations, and 60\% identified code walkthroughs as the most valuable mentorship aspect.

From the expert perspective, both reported spending 20--40 hours per new hire, with support frequency ranging from weekly to several times per week. Experts strongly agreed they repeatedly explain the same concepts (M=4.5/5), that switching to mentoring is cognitively exhausting (M=4.0/5), and that knowledge loss is a major organizational concern (M=4.0/5). Both identified missing documentation as a key challenge (M=5/5), and one expert noted many areas of the codebase are understood by only a few people.
These patterns, including informal mentorship, barriers to asking questions, heavy reliance on experts, repetitive explanations, and concentrated knowledge, directly motivated our evaluation.

\textbf{Study Design.} The controlled study employed a within-subjects design comparing two conditions: (1) Guided Tours, expert-prepared AI-assisted tours and accompanying podcast; and (2) Exploratory Tours, learner-generated AI-only tours without expert input. A traditional documentation baseline was not included as Bankhet lacked any existing documentation. Each learner completed both conditions in separate sessions of approximately 60 minutes, covering different features. We counterbalanced condition order and swapped which feature was used in each condition across participants to reduce learning effects and feature-difficulty confounds.

\textbf{Materials.} Expert A prepared two guided tours using the AI-assisted pipeline, one for each critical Bankhet operation of comparable complexity with 10-question quizzes and accompanying podcasts. To reduce evaluation bias, Expert B separately developed an evaluation rubric and rated all learner verbal explanations without knowledge of which condition each learner had completed.

\textbf{Procedure.} Sessions followed a standardized protocol: introduction, learning phase, comprehension quiz, and verbal explanation, evaluated using the rubric. Learners could freely navigate the repository and use supplementary tools. Screen recordings captured navigation patterns; the observer noted documented behaviors.

\textbf{Data Collection.} We collected quiz scores (0--100), expert ratings (1--5) for learner comprehension, and post-study survey responses (LS-Post, ES-Post), alongside qualitative data from recordings and observer notes. The full protocol and all instruments are available in our replication package.

\section{Case Study Findings} \label{sec: Case Study Findings}
We triangulate evidence from surveys, comprehension assessments, and semi-structured interviews. The replication package includes a summary of survey results across all 70 questions from surveys (LS-Pre, LS-Post, ES-Pre, ES-Post) and a question mapping for RQs\footnote{\url{https://figshare.com/s/6a261d3382b116d8494f}}.

\enlargethispage{\baselineskip}
\subsection{RQ1: Expert Support Effectiveness}

Lacy substantially reduced the expert burden established in Section~\ref{sec:case-study}. Both experts reported that creating tours was less time-consuming than having code walkthroughs with learners in meetings. Post-study ratings indicated decreased cognitive load (M=4.5/5) and less mental effort compared to manual documentation (M=4.5/5). Where pre-study data showed repetitive explaining as a major burden (M=4.5/5), experts agreed that Lacy reduces these repetitive aspects through AI-assisted tour generation (M=3.5/5), allowing them to focus on what AI cannot infer: domain context, design rationale, and organizational knowledge.

AI-generated content proved reliable as a starting point. Experts generally found tour descriptions closely related to the code files, and neither reported needing to correct inaccuracies. When edits were made, they focused on adjusting complexity (2/2), adding domain-specific context (1/2), and including warnings or edge cases (1/2). Similarly, experts considered AI-generated quizzes relevant to selected code sections and capable of testing learners' understanding of code logic, rarely requiring modification. Both rated tours as ``very effective'' starting points and felt confident learners could understand the intent of key code segments without constant real-time support (M=4.0/5).

Unlike ephemeral walkthroughs, tours create reusable knowledge assets where insights and tacit knowledge are documented once and shared many times. Both experts reported improved knowledge sharing (one ``major,'' one ``minor'' improvement), directly addressing pre-study knowledge concentration concerns. This shift fundamentally changes how expert knowledge scales,  addressing the challenge experts face with limited available time.

Both stated they would likely use Lacy for maintaining their projects going forward. One expert noted: ``onboarding has effectively been reduced to a couple of minutes of setup and guidance with code tours I prepared, instead of time-consuming walkthrough sessions.'' and emphasized ``the real differentiation for Lacy is the structured code walkthrough experience rather than generic AI explanations''. These findings suggest Lacy effectively supports experts by enabling less effortful knowledge capture that scales across learners without repeated real-time involvement.

\subsection{RQ2: Learner Support Effectiveness}
Lacy enabled effective comprehension of unfamiliar legacy code. During sessions, we observed participants relying on tours as their primary learning source, though some supplemented with GitHub Copilot for function-level detail (L2, L3, L4). One learner (L4) opted for GitHub Copilot to have a one-sentence explanation for each function defined in the code for an overview. Notably, two participants (L2, L4) had no prior exposure to our system; following the standard 10-minute introduction, both required no additional guidance and indicated intuitive design.

\begin{figure}
\centering
\begin{tikzpicture}
\begin{axis}[
    xbar,
    width=0.9\columnwidth,
    height=3.8cm,
    xlabel={Mean rating (1=Much Worse, 3=Equal, 5=Much Better)},
    xlabel style={font=\small},
    xmin=1, xmax=5,
    xtick={1,2,3,4,5},
    symbolic y coords={
        {Expert 1-on-1},
        {Docs},
        {Comments},
        {GPT/Copilot},
        {Code alone}
    },
    ytick=data,
    y tick label style={font=\small},
    bar width=8pt,
    nodes near coords,
    nodes near coords align={horizontal},
    every node near coord/.append style={font=\small\bfseries},
    enlarge y limits=0.12,
    extra x ticks={3},
    extra x tick style={grid=major, grid style={dashed, gray}},
]
\addplot[fill=blue!60, draw=blue!70] coordinates {
    (3.2,{Expert 1-on-1})
    (3.6,{Docs})
    (4.2,{Comments})
    (4.4,{GPT/Copilot})
    (4.6,{Code alone})
};
\end{axis}
\end{tikzpicture}
\caption{Lacy tours compared to traditional methods.}
\label{fig:comparison}
\end{figure}
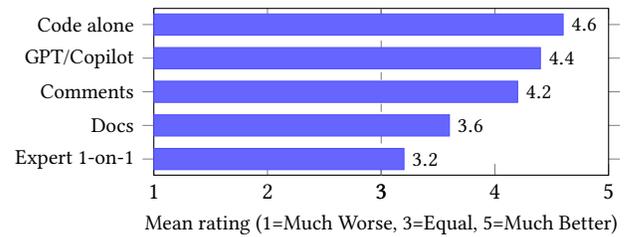

Central to Lacy's value is its ability to simulate expert mentoring. Learners agreed that both tour types felt like a senior developer walking them through code (M=4.1/5) and could effectively simulate expert guidance (Guided: M=4.2/5; Exploratory: M=3.8/5). Open-ended responses revealed what underlies this perception: \textit{structure}. Learners described being ``directed to the appropriate starting point,'' experiencing ``organized structure away from the chaos,'' and grasping ``the big picture''. This structured guidance on knowing where to start and what matters is precisely what experts naturally provide in walkthroughs and other methods lack. Learners rated Lacy favorably relative to traditional onboarding approaches (Figure~\ref{fig:comparison}), while 1-on-1 human expert walkthroughs were rated near parity. This positioning suggests that tours can approximate the expert walkthrough experience while remaining scalable, repeatable, and asynchronous. All learners rated Lacy as better than their prior onboarding experiences (60\% much better, 40\% somewhat better).

\begin{figure}[H]
\centering
\begin{tikzpicture}
\begin{axis}[
    xbar,
    width=0.85\columnwidth,
    height=3.5cm,
    xlabel={Score (\%)},
    xlabel style={font=\small},
    symbolic y coords={{Expert\\Rating}, {Quiz\\Score}},
    ytick=data,
    yticklabel style={font=\small, align=right, text width=1.2cm},
    xmin=0, xmax=100,
    bar width=12pt,
    legend style={
        at={(0.5,-0.2)}, 
        anchor=north, 
        legend columns=2, 
        font=\footnotesize,
        draw=none,
        column sep=0.3cm
    },
    area legend,
    enlarge y limits=0.4,
    nodes near coords,
    nodes near coords align={horizontal},
    every node near coord/.append style={font=\small\bfseries},
    extra x ticks={40},
    extra x tick style={grid=major, grid style={dashed, gray, thick}},
    extra x tick labels={},
]
\addplot[fill=blue!60, draw=blue!70] coordinates {
    (83,{Quiz\\Score})
    (79,{Expert\\Rating})
};
\addplot[fill=orange!60, draw=orange!70] coordinates {
    (57,{Quiz\\Score})
    (76.8,{Expert\\Rating})
};

\addlegendimage{area legend, fill=blue!60, draw=blue!70}
\addlegendimage{area legend, fill=orange!60, draw=orange!70}
\legend{Guided, Exploratory}
\end{axis}
\end{tikzpicture}
\caption{Comprehension outcomes by condition. The dashed line indicates the expected baseline.}
\label{fig:comprehension}
\end{figure}
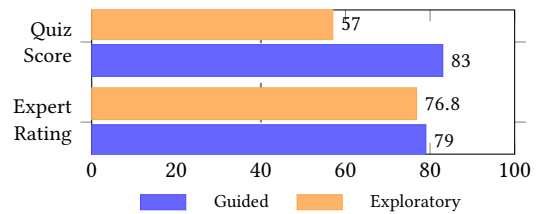

Comprehension data revealed that expert curation adds substantial value beyond AI-only generation, even when learners do not fully perceive it. Guided tours yielded +26 percentage points on quizzes (83\% vs 57\%) and higher expert ratings (79\% vs 76.8\%), despite being 3--5 steps longer (Figure~\ref{fig:comprehension}). Participants completed guided tours approximately 10 minutes faster ($\sim$25 vs $\sim$35 min), suggesting expert curation produces more focused, efficient learning paths. Interestingly, learners only moderately agreed that expert curation combined with AI was more valuable than AI alone (M=3.4/5) however, their performance told a different story. This perception-performance gap may partly reflect how AI-only tours produce smooth explanations that feel immediately intuitive, creating a sense of understanding. This may also suggest an over-reliance on AI, where learners trust AI-generated content as sufficient without recognizing what is missing. Organizationally, developers cannot reliably self-assess whether expert curation is needed, meaning organizations that rely on satisfaction ratings alone will systematically undervalue it and default to the faster AI-only path. That both tour types were perceived as effectively simulating expert guidance (Guided M=4.2/5, Exploratory M=3.8/5). Nevertheless, all participants preferred using both tour types together in an ideal onboarding scenario, recognizing their complementary roles. Additionally, the podcast feature received moderate ratings across helpfulness, engagement, and mentor simulation (M=3.2--3.4/5), and we believe it holds potential as a stepping stone toward richer multimodal learning environments for software onboarding

Beyond comprehension scores, learners demonstrated confidence in their acquired knowledge. They agreed they felt confident explaining what they learned (M=4.2/5), had learned real-world tips and best practices (M=4.0/5), and viewed tours as an effective starting point for understanding the codebase (M=4.4/5). Aggregating learning outcome metrics (understood faster, learned effectively, effective starting point) yields a mean of 4.5/5, indicating strong perceived learning efficacy. Critically, these outcomes translated to reduced expert dependency. All learners agreed they would ask fewer questions after completing tours (M=5.0/5) and that well-made tours could significantly reduce time with experts (M=4.8/5). These findings suggest Lacy enables self-directed onboarding by incorparating expert value.

\subsection{RQ3: Organizational Integration}
Interviews with stakeholders revealed that knowledge concentration poses significant organizational risk and reduces productivity. One expert reported spending approximately one day per week in onboarding meetings, suggesting a substantial productivity impact. Interviewees agreed that code tours and the assignment system can address onboarding challenges by mimicking expert walkthroughs within a systematic workflow: experts create tours once, assign them to learners, and track completion through a centralized dashboard. Stakeholders confirmed that this workflow transforms project knowledge into a managed organizational asset rather than scattered tribal knowledge.

The reusability of code tours translates to quantifiable impact. Creating a guided tour required approximately 30 minutes of expert time, compared to 20 minutes for a typical live walkthrough. Unlike walkthroughs, tours can onboard multiple learners without repeated expert involvement. If a tour is used by $N$ learners, the per-learner expert time drops from 20 minutes to $\frac{30}{N}$ minutes— substantial time and cost savings when onboarding newcomers.

Stakeholders emphasized that preserving business flow is Beko's main priority in project maintainability. During interviews, an expert noted Lacy ``has the potential to preserve business logic and design choices'', directly addressing the bus factor concern. One senior developer suggested that existing code tours could enable faster response in crisis situations requiring quick action on unfamiliar code, without waiting for expert availability. Additionally, stakeholders discussed potential value for short-term internships, where tours could transfer only necessary knowledge efficiently to maximize productivity within limited timeframes.

All stakeholders agreed on Lacy's workflow integration and intuitive design. Experts could envision Lacy fitting into existing workflows (M=4.0/5), found the learning curve minimal (M=3.5/5), and reported that updating tours was easier than traditional support methods (M=4.0/5). Following these findings, Beko deployed Lacy for onboarding, providing evidence that it can scale to organizational needs.

\section{Lessons Learned}
Based on our deployment and evaluation, we identify the following implications for practitioners designing AI-assisted onboarding tools. These lessons extend beyond the requirements identified in Table~\ref{tab:requirements} to broader, transferable design principles. 

\textbf{Use code tours as situated, expert-like guidance.} Code tours are an intuitive mechanism for communicating industry practices because they resemble an expert walking a newcomer through a codebase. Being IDE-integrated, tied to concrete code locations, and structured as sequential narratives, they provide explanations that better support onboarding than disconnected documentation.

\textbf{Embrace hybrid human--AI authoring.} Fully AI-generated documentation can miss tacit knowledge, while purely manual approaches over-rely on time-constrained experts. Effective onboarding tools should be hybrid: experts curate what matters (key components, business logic, intent), while AI operationalizes this guidance (drafting, structuring, linking), improving quality without increasing documentation overhead. 

\textbf{Capture knowledge at natural moments of explanation.} Rather than requiring dedicated documentation efforts, expert knowledge is easiest to capture when it is already being expressed. Tools should capture natural information transfer during daily practice such as recording experts as they highlight code and explain it during walkthroughs, then structuring this content for later reuse and sharing through artifacts such as our voice-to-tour feature. This approach can lower documentation friction and facilitate the capture of tacit knowledge.

\section{Threats to Validity} \label{sec: Threats to Validity}
The success of AI-assisted guided tours and exploratory tours is dependent on LLM success and expert input (e.g., selected code snippets or files, and provided insights) and the project and codebase itself. Variations in these factors may impact the quality and accuracy of the results. Some organizations may reject external LLM usage due to security or privacy reasons. Lacy could instead use alternatives such as internal LLMs or local models in such settings.

Our data collection is based on self-reported surveys and interviews, and may be subjective, and their results may differ with a larger or different participant pool. We recruited the two experts and the five developers actively assigned to Bankhet tasks to reflect the project’s real mentoring and learning context.

Because tours reference concrete code artifacts, they may become stale as the codebase evolves. However, unlike standalone documentation, tours are lightweight and tied to code navigation, making incremental updates more natural than maintaining a separate knowledge base. Since tours are also quick to create, they can be reconstructed on demand as needs arise, making staleness less of a systemic concern. While our mature legacy setting reduced this risk, staleness remains relevant for more actively developed systems and warrants attention in future work.

\section{Conclusion}
In this study, we introduced Lacy, an onboarding system that utilizes code tours to simulate expert mentoring by providing step-by-step explanations inside the IDE, highlighting relevant code snippets across and within files to ease navigation. Over the course of a year, we collaborated with our industry partner Beko to identify onboarding challenges, derive requirements (Table~\ref{tab:requirements}), design and improve the system. We deployed the system, and further conducted a case study on a legacy codebase, systematically assessing how Lacy supports experts, learners, and the organization. Results showed that Lacy reduced expert burden through reusable tours, improved learner comprehension compared to AI-only approaches, and integrated seamlessly into existing workflows. Notably, all stakeholders found the system intuitive, and Beko has since adopted Lacy for onboarding use. To our knowledge, Lacy represents the first hybrid human-AI approach to code tour creation. Finally, we release our code and case study instruments in the replication package. We believe that code tour-based interactions have broader implications for knowledge transfer, software engineering education, and software documentation beyond onboarding. Future work includes generating code tours from codebase analysis, version control, extracting historical data or design documents, and proposing starter implementation tasks for newcomers beyond quizzes.

\newpage
\section*{Acknowledgment} \label{sec:Acknowledgment}
We thank the developers and stakeholders at Beko for their feedback. We acknowledge the use of AI-assisted tools for grammar and language editing. This work has been supported by the ITEA4 GENIUS project, funded by the national funding authorities of the participating countries: https://itea4.org/project/genius.html

\enlargethispage{4\baselineskip}
\bibliographystyle{ACM-Reference-Format}
\bibliography{references}
\end{document}